\documentclass[11pt]{article}
\newlength{\vshift}
\newlength{\hshift}
\usepackage{amssymb,amsopn}

\def\de{\delta}

\def\ds{\stackrel{\star}{,}}

\def\p{\partial}

\def\hp{\hat{\partial}}

\def\h{\hat}
\def\lb{\lbrack}
\def\rb{\rbrack}

\begin{document}

\begin{titlepage}
\rightline{MPP-2004-111}

\vspace{2em}
\begin{center}

{\Large{\bf Deformed Bialgebra of Diffeomorphisms}}

\vskip 3em

{{\bf Marija Dimitrijevi\' c${}^{2,3}$, Julius Wess${}^{1,2}$ }}

\vskip 1em

${}^{1}$Universit\"at M\"unchen, Fakult\"at f\"ur Physik\\
        Theresienstr.\ 37, 80333 M\"unchen, Germany\\[1em]

${}^{2}$Max-Planck-Institut f\"ur Physik\\
        F\"ohringer Ring 6, 80805 M\"unchen, Germany\\[1em]

${}^{3}$University of Belgrade, Faculty of Physics\\
Studentski trg 12, 11000 Beograd, Serbia and Montenegro\\[1em]
\end{center}

\vspace{2em}

\begin{abstract}
The algebra of diffeomorphisms derived from general coordinate transformations on commuting coordinates
is represented by differential operators on noncommutative spaces. The algebra remains unchanged,
the comultiplication however is deformed, that way we have found a deformed bialgebra of diffeomorphisms.
Scalar, vector and tensor fields are defined with appropriate transformation laws under the
deformed algebra and a differential calculus is developed. For pedagogical reasons the formalism is
developed for the $\theta$-deformed space as it is the best known example of deformed spaces.
\end{abstract}
\vspace*{0.5cm}
\begin{center}
Talk given by Marija Dimitrijevi\' c{\footnote{see Introduction}} at the 1st Vienna Central European Seminar
on\\
\hspace*{-2.5cm}Particle Physics and Quantum Field Theory, 26-28 November 2004
\end{center}
\vspace*{0.55cm}


\hspace*{0.6cm}MSC: 81T75 Noncommutative geometry methods, 58B22 Geometry of Quantum groups\\

\vspace*{0.5cm}
\quad\scriptsize{eMail: dmarija,wess@theorie.physik.uni-muenchen.de}
\vfill

\end{titlepage}\vskip.2cm

\newpage
\setcounter{page}{1}

\section{Introduction}
This lecture is based on common work with P. Aschieri, C. Blohmann, F. Meyer amd P. Schupp
that will be published in a forthcomming paper \cite{1}.

Deformed coordinate spaces based on algebraic relations have been studied extensively \cite{2}. The
$\theta$-, $\kappa$- and $q$-deformations are the best known examples \cite{3}. In this lecture we shall treat the
$\theta$-deformed space for simplicity, the formalism however is presented in such a way that the generalisation
to other spaces is obvious.

It is not clear how coordinate transformations can be defined on noncommuting coordinates
such that this leads to a meaningful calculus. Our strategy is as follows: we start from general coordinate transformations
of commuting variables. Their algebraic structure is represented by algebra of diffeomorphisms, a Lie
algebra with well defined comultiplicaton rules.  We shall deforme their structure and take this as the basis
for a deformed differential geometry. Our approach makes heavy use of the $\star$-product
formalism \cite{4}. In the second chapter we show how to derive the $\star$-product from the algebra \cite{5}.
For the $\theta$-deformed space we arrive at the Moyal-Weyl product, as expected. The properties of the Moyal-Weyl
product are well known, nevertheless we go trough a rather detailed analysis having a generalisation to
other quantum spaces in mind.

As a next step we define derivatives. A differential calculus on noncommuting spaces has been
established in previous work \cite{6}, we only have to show how that calculus can be expressed in the $\star$-product
formalism. Vector fields on noncommuting spaces are represented in
the $\star$-product formalism as higher order differential
operators acting on functions of commuting variables. This suggests to study higher order differential operators
on the deformed and nondeformed spaces as well. It turns out that there is an invertible map from the differential
operator on commuting variables to differential operators on noncommuting variables preserving the algebraic
structure. When we apply
this to the vector fields on commuting variables we obtain higher order differential operators on
noncommuting spaces that form the same algebra as the original vector fields. That way we found a representation
of the algebra of diffeomorphisms as higher order differential operators on noncommuting spaces.
The algebra remains unchanged, the comultiplication changes. We have constructed a deformed bialgebra of diffeomorphisms.

In the last chaper we show how this algebra acts on scalar field and on covariant and contravariant vector
fields. This should pave the way to a deformed differential geometry on noncommutative spaces that finally leads to
a deformed gravity theory.

\section{Properties of the deformed coordinate spaces}

The deformed coordinate spaces $\hat{\cal{A}}_{\hat{x}}$ that we have
been considering, the $\theta$-, $\kappa$- and $q$-deformed spaces, are
based on coordinates $\hat{x}=\{\hat{x}^1,\dots,\hat{x}^d\}$ and on the respective relations
\cite{7}. The coordinates form
an associative free algebra
and $\hat{\cal{A}}_{\hat{x}}$ is the quotient
of this algebra by the ideal generated by the relations \cite{8}.
The spaces mentioned above have additional properties that are useful for
developing a differential calculus, a star product formulation and for
application in physics. We mention two of them here.

\subsection{Conjugation}

It is possible to define a conjugation on the coordinates \cite{9}, $\bar{\hat{x}}^\nu$ such that
$\bar{i}=-i$ and $\overline{\hat{x}^\mu\hat{x}^\nu}=\bar{\hat{x}}^\nu\bar{\hat{x}}^\mu$
and such that the relations are left invariant. In this case we can
identify:
\begin{equation}
\bar{\hat{x}}^\nu=\hat{x}^\nu \label{1}
\end{equation}
and define real coordinates.

For the $\theta$-deformed space, this is the space we are going to
deal with in this lecture, we have the relation
\begin{equation}
\lb \h{x}^\mu , \h{x}^\nu\rb = i \theta^{\mu\nu}, \label{2}
\end{equation}
and consequently, for real $\theta$
\begin{equation}
\lb \bar{\h{x}}^\mu , \bar{\h{x}}^\nu\rb = i \theta^{\mu\nu}, \label{3}
\end{equation}
and thus equation (\ref{1}) can be imposed.

\subsection{Poincar\' e-Birkhoff-Witt (PBW) property}

This concept was first developed for Lie algebras \cite{10}, but it applies to
the other deformed algebras as well. PBW demands that the dimensions of the finite
dimensional vector spaces, spanned by the homogeneous polynomials of
degree $r$, have the same dimension as the corresponding vector spaces of
commuting variables. We shall denote these vector spaces $\hat{V}_r$
and $V_r$, respectively. PBW gives some information on the ``size'' of the
algebra $\hat{\cal{A}}_{\hat{x}}$ which is infinite dimensional. The
$\theta$-, $\kappa$- and $q$-deformed spaces have PBW property.

\section{The algebra and the star product ($\star$-product)}

Finite dimensional vector spaces with the same dimension are isomorphic:
\begin{equation}
\hat{V}_r\sim V_r .\label{4}
\end{equation}
To establish this isomorphism we choose a basis in $\hat{V}_r$ and map
this basis on a suitable basis in $V_r$. The specific form of the isomorphism
will depend on the basis chosen. A natural choice are the completely
symmetrisized monomials. We denote the elements of this basis with $:\>:$
\begin{eqnarray}
:\hat{x}^\mu:&=&\hat{x}^\mu, \nonumber\\
:\hat{x}^\mu\hat{x}^\nu:&=&\frac{1}{2}(\hat{x}^\mu\hat{x}^\nu+\hat{x}^\nu\hat{x}^\mu),\quad {\mbox{ etc. }}\label{5}
\end{eqnarray}
The vector space $\hat{V}=\sum_r \oplus\hat{V}_r$ is spanned by this basis, $V$ is the
corresponding vector space of commuting variables.

The vector space isomorphism $\varphi$ is then defined by a map of the basis in the
noncommutative spaces on the corresponding
basis of commuting variables:
\begin{eqnarray}
\varphi:\>\hat{V}\rightarrow V \quad\quad
\varphi: &C_0&\hspace*{-0.3cm}+C_{1\mu}:\hat{x}^\mu:+C_{2\mu\nu}:\hat{x}^\mu\hat{x}^\nu:+\dots \nonumber\\
\mapsto \hspace*{-0.3cm}&C_0&\hspace*{-0.3cm}+C_{1\mu}x^\mu+C_{2\mu\nu}x^\mu x^\nu +\dots .\label{6}
\end{eqnarray}
This is a vector space isomorphism with the inverse
\begin{eqnarray}
\varphi^{-1}: &C_0&\hspace*{-0.3cm}+C_{1\mu}x^\mu+C_{2\mu\nu}x^\mu x^\nu +\dots \nonumber\\
\mapsto \hspace*{-0.3cm}&C_0&\hspace*{-0.3cm}+C_{1\mu}:\hat{x}^\mu:
+C_{2\mu\nu}:\hat{x}^\mu\hat{x}^\nu:+\dots .\label{7}
\end{eqnarray}
As an example consider the element $\hat{x}^\mu\hat{x}^\nu$, it has to be expanded in the basis:
\begin{eqnarray}
\hat{x}^\mu\hat{x}^\nu\hspace*{-0.3cm}&=&\hspace*{-0.3cm}\frac{1}{2}(\hat{x}^\mu\hat{x}^\nu
+\hat{x}^\nu\hat{x}^\mu)+\frac{1}{2}(\hat{x}^\mu\hat{x}^\nu-\hat{x}^\nu\hat{x}^\mu) \nonumber\\
&=&\hspace*{-0.3cm}:\hat{x}^\mu\hat{x}^\nu:+\frac{i}{2}\theta^{\mu\nu}. \label{8}
\end{eqnarray}
By $\varphi$, this element is mapped as follows:
\begin{equation}
\varphi:\quad \hat{x}^\mu\hat{x}^\nu \mapsto x^\mu x^\nu+\frac{i}{2}\theta^{\mu\nu}. \label{9}
\end{equation}
The same procedure leads to
\begin{equation}
\varphi:\quad \hat{x}^\nu\hat{x}^\mu \mapsto x^\mu x^\nu-\frac{i}{2}\theta^{\mu\nu}. \label{10}
\end{equation}
Thus,
$$
\varphi:\quad \hat{x}^\mu\hat{x}^\nu-\hat{x}^\nu\hat{x}^\mu \mapsto i\theta^{\mu\nu}.
$$
The inverse isomorphism  $\varphi^-1$ is straightforward
\begin{equation}
\varphi^{-1}:\quad x^\mu x^\nu\mapsto :\hat{x}^\mu\hat{x}^\nu:=
\frac{1}{2}(\hat{x}^\mu\hat{x}^\nu+\hat{x}^\nu\hat{x}^\mu) .\label{11}
\end{equation}

The vector space isomorphism $\varphi$ can be extended to an algebra morphism. We start
with two elements of $\hat{\cal{A}}_{\hat{x}} $:
\begin{equation}
\hat{g}(\hat{x})\in \hat{\cal{A}}_{\hat{x}} ,\quad \hat{f}(\hat{x})\in
\hat{\cal{A}}_{\hat{x}} . \label{12}
\end{equation}
The product is well-defined and again an element of $\hat{\cal{A}}_{\hat{x}} $:
\begin{equation}
\hat{f}(\hat{x})\hat{g}(\hat{x})= {\hat{f}}\cdot{\hat{g}} (\hat{x})\in \hat{\cal{A}}_{\hat{x}} .\label{13}
\end{equation}
The elements of $\hat{\cal{A}}_{\hat{x}}$ can be expanded in
the basis chosen and mapped by $\varphi$ on the algebra of commuting variables
${\cal{A}}_x$:
\begin{eqnarray}
\varphi:\> &\hat{f}&\hspace*{-0.3cm}(\hat{x}) \mapsto f(x)\in {\cal{A}}_x ,\nonumber\\
&\hat{g}&\hspace*{-0.3cm}(\hat{x})\mapsto g(x)\in {\cal{A}}_x , \label{14}\\
&\hat{f}&\hspace*{-0.35cm}\cdot\hspace*{0.5mm}{\hat{g}}({\hat{x}})\mapsto f\star g(x) \in {\cal{A}}_x .
\nonumber
\end{eqnarray}
This defines the $\star$-product of two functions. The algebra of the noncommuting variables,
$\hat{\cal{A}}_{\hat{x}}$ is isomorphic to the algebra of commuting variables with
the $\star$-product as multiplication. As an example we consider first the elements
$\hat{f}({\hat{x}})=\hat{x}^\mu $ and
$\hat{g}({\hat{x}})=\hat{x}^\nu $.
From (\ref{5}) follows $f=x^\mu$ and $g=x^\nu$ and (\ref{9}) and (\ref{10}) yield:
\begin{eqnarray}
x^\mu\star x^\nu \hspace*{-0.3cm}&=&\hspace*{-0.3cm}x^\mu x^\nu +\frac{i}{2}\theta^{\mu\nu},\nonumber\\
x^\nu\star x^\mu \hspace*{-0.3cm}&=&\hspace*{-0.3cm}x^\mu x^\nu -\frac{i}{2}\theta^{\mu\nu} .\label{15}
\end{eqnarray}
This result can be written with a bidifferential operator
\begin{eqnarray}
x^\mu\star x^\nu &=& x^\mu x^\nu + \frac{i}{2}\theta^{\rho\sigma}\partial_\rho x^\mu
\partial_\sigma x^\nu \nonumber\\
&=&\sum _{n=1}^\infty \Big(\frac{i}{2}\Big)^n \frac{1}{n!}\theta^{\rho_1\sigma_1}\dots
\theta^{\rho_n\sigma_n}\Big(\partial_{\rho_1}\dots\partial_{\rho_n}x^\mu\Big)
\Big(\partial_{\sigma_1}\dots\partial_{\sigma_n}x^\nu\Big) .\label{16}
\end{eqnarray}
This is easy to see because the higher derivatives vanish. We shall show that the
$\star$-product of two elements of ${\cal{A}}_x$ can always be written in this form for arbitrary polynomials
$f$ and $g$.
\begin{equation}
f\star g\,(x)=  \sum _{n=1}^\infty \Big(\frac{i}{2}\Big)^n \frac{1}{n!}\theta^{\rho_1\sigma_1}\dots
\theta^{\rho_n\sigma_n}\Big(\partial_{\rho_1}\dots\partial_{\rho_n}f(x)\Big)
\Big(\partial_{\sigma_1}\dots\partial_{\sigma_n}g(x)\Big) .\label{17}
\end{equation}
This way the $\star$-product of two polynomials can be formally extended to the $\star$-
product of two smooth functions $f$ and $g$; $f\in \mathrm{C}^\infty$,
$g\in \mathrm{C}^\infty$.

We shall prove (\ref{17}) by induction on the degree of the polynomial $\hat{f}(\hat{x})$.
For $\hat{f} = 1$  the statement is obvious. We assume that it is true
for $\hat{f}$ of degree $r$ and prove it for polynomials of degree $r+1$. To
proceed we first prove it for $\hat{f}$ linear acting on elements of the basis:
\begin{equation}
\hat{x}^\rho :\hat{x}^{\mu_1}\dots \hat{x}^{\mu_n}:=
\hat{x}^\rho\frac{1}{n!}\sum _{P\{\mu_1\dots\mu_n\}}\hat{x}^{\mu_1}\dots \hat{x}^{\mu_n}. \label{18}
\end{equation}

With the relation (\ref{2}) we can take $\hat{x}^\rho$ to any position
in this product. We add up all terms generated that way to obtain a polynomial
symmetric in $\rho ,\mu_1,\dots \mu_n$:            :
\begin{eqnarray}
(n+1)\hat{x}^\rho :\hat{x}^{\mu_1}\dots \hat{x}^{\mu_n}: &=& \frac{1}{n!}
\sum _{P\{\rho,\mu_1\dots\mu_n\}}\hat{x}^\rho\hat{x}^{\mu_1}\dots \hat{x}^{\mu_n} \nonumber\\
&+& \hspace*{-0.3cm}\frac{1}{n!}\frac{n(n+1)}{2}\sum_{P\{\nu_1\dots\nu_n\}}\frac{i}{2}\theta^{\rho\nu_j}
\big(\hat{x}^{\nu_1}\dots \hat{x}^{\nu_n}\big)\Big|_{\mbox{\footnotesize{$\nu_j$\hspace*{0.7mm}missing}}},
\label{19}
\end{eqnarray}
or
\begin{equation}
\hat{x}^\rho :\hat{x}^{\mu_1}\dots \hat{x}^{\mu_n}:\>=\>:\hat{x}^\rho\hat{x}^{\mu_1}\dots \hat{x}^{\mu_n}:
+\frac{i}{2}\sum _j \theta^{\rho\nu_j}
:\hat{x}^{\nu_1}\dots \hat{x}^{\nu_n}:\Big|_{\mbox{\footnotesize{$\nu_j$\hspace*{0.7mm}missing}}}. \label{20}
\end{equation}
This is exactly the result that we obtain from (\ref{17}). The rest of the proof
follows from associativity:
\begin{equation}
(f\star g)\star h= f\star (g\star h).\label{21}
\end{equation}
For the associative algebra associativity holds by definition, for the
$\star$-product (\ref{17}) it has to be shown explicitly, as we do not know that the $\star$-product (\ref{17})
really is the right $\star$-product. We are in the middle of the proof. We use a
specific notation for (\ref{17})
\begin{equation}
f\star g\, (x)=e^{\frac{i}{2}\frac{\p }{\p x^\rho}\theta^{\rho\sigma}\frac{\p }{\p y^\sigma}}f(x)g(y)
\Big|_{y\rightarrow x}.\label{22}
\end{equation}
In this notation the following identity holds:
\begin{eqnarray}
\p _x \Big(f(x)g(x)\Big) &=& \frac{\p}{\p x}\Big(f(y)g(z)\Big|_{y,z\rightarrow x}\Big) \nonumber\\
&=&\Big(\frac{\p}{\p y}+\frac{\p}{\p z}\Big)f(y)g(z)\Big|_{y,z\rightarrow x}. \label{23}
\end{eqnarray}
This is the Leibniz rule. It also holds for any power of the derivative and,
therefore, for any polynomial $P(\frac{\p}{\p x})$
\begin{equation}
P(\frac{\p}{\p x})\Big( \big( f(y)g(z)\big) \Big|_{y,z\rightarrow x}\hspace*{0.1cm}\Big)
=P(\frac{\p}{\p y}+\frac{\p}{\p z})
f(y)g(z)\Big|_{y,z\rightarrow x}. \label{24}
\end{equation}
We rewrite the left-hand side of (\ref{22}) in this notation
\begin{eqnarray}
(f\star g)\star h\,(x) &=& e^{\frac{i}{2}\frac{\p }{\p x^\rho}\theta^{\rho\sigma}
\frac{\p }{\p z^\sigma}}(f\star g)(x)h(z)\Big|_{z\rightarrow x} \nonumber\\
&=&e^{\frac{i}{2}\frac{\p }{\p x^\rho}\theta^{\rho\sigma}
\frac{\p }{\p z^\sigma}}\Big( \big( e^{\frac{i}{2}\frac{\p }{\p x^\mu}\theta^{\mu\nu}
\frac{\p }{\p y^\nu}}f(x)g(y)\big)\Big|_{y\rightarrow x}h(z)\Big)\Big|_{z\rightarrow x}\nonumber\\
&=& e^{\frac{i}{2}(\frac{\p }{\p x^\rho}+\frac{\p }{\p y^\rho})\theta^{\rho\sigma}
\frac{\p }{\p z^\sigma}}e^{\frac{i}{2}\frac{\p }{\p x^\mu}\theta^{\mu\nu}
\frac{\p }{\p y^\nu}}f(x)g(y)h(z)\Big|_{y,z\rightarrow x} \label{25}\\
&=& e^{\frac{i}{2}\big(\frac{\p }{\p x^\rho}\theta^{\rho\sigma}
\frac{\p }{\p z^\sigma}+\frac{\p }{\p y^\rho}\theta^{\rho\sigma}
\frac{\p }{\p z^\sigma}+\frac{\p }{\p x^\rho}\theta^{\rho\sigma}
\frac{\p }{\p y^\sigma}}f(x)g(y)h(z)\Big|_{y,z\rightarrow x}. \nonumber
\end{eqnarray}
Analogously
\begin{eqnarray}
f\star (g\star h)\,(x) &=& e^{\frac{i}{2}\frac{\p }{\p x^\rho}\theta^{\rho\sigma}
\frac{\p }{\p y^\sigma}}f(x)\big( g\star h\big) (y)\Big|_{y\rightarrow x} \nonumber\\
&=& e^{\frac{i}{2}\frac{\p }{\p x^\rho}\theta^{\rho\sigma}
\frac{\p }{\p y^\sigma}}f(x)\Big( e^{\frac{i}{2}\frac{\p }{\p y^\mu}\theta^{\mu\nu}
\frac{\p }{\p z^\nu}}g(y)h(z)\Big|_{z\rightarrow y}\Big) \label{26}\\
&=& e^{\frac{i}{2}\big(\frac{\p }{\p x^\rho}\theta^{\rho\sigma}
(\frac{\p }{\p y^\sigma}+\frac{\p }{\p z^\sigma})
+\frac{\p }{\p y^\rho}\theta^{\rho\sigma}\frac{\p }{\p z^\sigma}\big)}f(x)g(y)h(z)
\Big|_{y,z\rightarrow x}. \nonumber
\end{eqnarray}
The two expressions agree, associativity is shown. For the rest
of the proof we choose in equation (\ref{22}) $x^\rho$ for $f$, a
polynomial of degree $r$ for $g$. The right-hand side is true by the
assumption of the induction, on the left-hand side $f\star g$ is a polynomial of degree $r+1$.

The $\star$-product (\ref{17}) was first introduced by H. Weyl \cite{11} for his
quantization procedure and later by Moyal \cite{12} to relate functions of the phase
space to quantum-mechanical operators in Hilbert space. For this reason the
$\star$-product (\ref{17}) occurs in the literature as Moyal-Weyl product.
Other $\star$-products will occur for different orderings.

The reality property is easily seen:
\begin{equation}
\overline{f\star g}=\bar{g}\star \bar{f} ,\label{27}
\end{equation}
where $\bar{f}$ is the complex conjugate of $f$ and $\bar{g}$ is the complex conjugate of $g$.

The $\star$-product (\ref{22}) is a bilinear map
\begin{eqnarray}
\star :\quad {\cal{A}}_x\times{\cal{A}}_x &\rightarrow & {\cal{A}}_x \nonumber\\
(f(x),g(x)) &\mapsto & f\star g (x).\label{28}
\end{eqnarray}
It defines a unique map $\tilde{\star}$ on the tensor product (universal mapping property of the
tensor product):
\begin{equation}
\tilde{\star} :\quad f\otimes g \mapsto e^{\frac{i}{2}\theta^{\rho\sigma}\partial _\rho
\otimes\partial_\sigma}f\otimes g \label{29}
\end{equation}
such that $\mu\cdot \tilde{\star}=\star$, that is
\begin{equation}
\mu\Big( e^{\frac{i}{2}\theta^{\rho\sigma}\partial _\rho
\otimes\partial_\sigma}f\otimes g \Big)= f\star g , \label{30}
\end{equation}
where $\mu$ maps $f\otimes g$ to $fg$. Many calculations become simpler if we
use the $\star$-product in the form (\ref{29}) and (\ref{30}).

\section{Derivatives and higher order differential operators}

\subsection{Derivatives}

Derivatives are defined as maps on $\hat{\cal{A}}_{\hat{x}}$ \cite{6,7}:
\begin{eqnarray}
\hat{\partial}: \quad \hat{\cal{A}}_{\hat{x}}&\rightarrow & \hat{\cal{A}}_{\hat{x}} \nonumber\\
\hat{f}(\hat{x}) &\mapsto &(\hat{\partial}\hat{f})(\hat{x}) .\label{31}
\end{eqnarray}
For the $\theta$-space a natural choice, compatible with the relation (\ref{2}),
is
\begin{equation}
\lb\hat{\partial}_\mu,\hat{x}^\nu\rb=\de_\mu^\nu ,\quad\quad
\lb\hat{\partial}_\mu, \hat{\partial}_\nu \rb =0. \label{32}
\end{equation}
This implies the usual Leibniz rule:
\begin{equation}
\hp_\mu \big( \hat{f}\hat{g}\big) = (\hp_\mu \hat{f})\hat{g} + \hat{f}(\hp_\mu \hat{g}). \label{33}
\end{equation}

By the isomorphism $\varphi$ the derivatives can be mapped to the space of
functions of commuting variables:
\begin{eqnarray}
\hat{f}(\hat{x})&\stackrel{\varphi}{\longmapsto}& f(x)\nonumber\\
\hat{\partial}\downarrow &&\downarrow\partial\star \label{34}\\
(\hat{\partial}\hat{f})(\hat{x})&\stackrel{\varphi}{\longmapsto}& (\partial\star f)(x).\nonumber
\end{eqnarray}
The map $\partial\star$ is defined by this diagram. For the derivatives
above we find
\begin{equation}
(\partial_\rho\star f)(x)=\partial_\rho f(x). \label{35}
\end{equation}
The derivatives act as the usual derivatives. This had to be expected
because the $\star$-operation is $x$ independent and thus commutes with the derivatives.

\subsection{Vector fields}

Vector fields can be defined on the algebra:
\begin{equation}
\hat{\xi}: \quad \hat{f}(\hat{x})\mapsto \hat{\xi}^\rho\hat{\partial}_\rho \hat{f}. \label{36}
\end{equation}
They map elements of $\hat{\cal{A}}_{\hat{x}}$ on elements of $\hat{\cal{A}}_{\hat{x}}$.
These, in turn, can be mapped to the function space of commuting variables:
\begin{equation}
\varphi: \quad  \hat{\xi}^\rho\hat{\partial}_\rho \hat{f} \mapsto
\xi^\rho\star (\partial_\rho\star)f=\xi^\rho\star\partial_\rho f ,\label{37}
\end{equation}
where
\begin{equation}
\varphi:\quad \hat{\xi}\mapsto\xi {\mbox{ and }} \hat{f}\mapsto f. \label{38}
\end{equation}
The vector field on $\hat{\cal{A}}_{\hat{x}}$ becomes a higher order
differential operator acting on $f$, if we insert the definition of a star product into (\ref{37}):
\begin{equation}
\xi^\rho\star\partial_\rho f=\sum_{n=0}^\infty \Big(\frac{i}{2}\Big)^n\frac{1}{n!}
\theta^{\mu_1\nu_1}\dots\theta^{\mu_n\nu_n}(\partial_{\mu_1}\dots\partial_{\mu_n}\xi^\rho)
\partial_{\nu_1}\dots\partial_{\nu_n}\partial_{\rho}f .\label{39}
\end{equation}
This suggests to study higher order differential
operators on the algebra $\hat{\cal{A}}_{\hat{x}}$ from the very beginning.

\subsection{Higher order differential operators}

Let $X$ be a higher order differential operator acting on
smooth function $f$ in ${\cal{A}}_x$:
\begin{eqnarray}
f\hspace*{-3mm}&\in &\hspace*{-3mm}{\cal{A}}_x \nonumber\\
Xf\hspace*{-3mm}&=&\hspace*{-3mm}\sum_r\xi_r^{\mu_1\dots\mu_r}
\partial_{\mu_1}\dots\partial_{\mu_r}f \in{\cal{A}}_x .\label{40}
\end{eqnarray}
In the same spirit $\hat{X}$ is a higher order differential operator
acting on elements of $\hat{\cal{A}}_{\hat{x}}$
\begin{eqnarray}
\hat{f}\hspace*{-3mm}&\in &\hspace*{-3mm}\hat{\cal{A}}_{\hat{x}} \nonumber\\
\hat{X}\hat{f}\hspace*{-3mm}&=&\hspace*{-3mm}\sum_r\hat{\xi}_r^{\mu_1\dots\mu_r}
\hat{\partial}_{\mu_1}\dots\hat{\partial}_{\mu_r}\hat{f} \in\hat{\cal{A}}_{\hat{x}} .\label{41}
\end{eqnarray}
The homomorphism $\varphi$ maps an element $\hat{X}\hat{f}$
of $\hat{\cal{A}}_{\hat{x}}$ to an element $\Xi f$ of ${\cal{A}}_x$:
\begin{eqnarray}
\varphi :\quad \hat{X}\hat{f}\hspace*{-3mm}&\mapsto &
\hspace*{-3mm}X\star f=\Xi _X f \in{\cal{A}}_x \nonumber\\
X\hspace*{-3mm}&=&\hspace*{-3mm}\sum_r\xi_r^{\mu_1\dots\mu_r}
\partial_{\mu_1}\dots\partial_{\mu_r}, \label{42}\\
\Xi_X\hspace*{-3mm}&=&\hspace*{-3mm}\sum_r\sum_n
\Big(\frac{i}{2}\Big)^n\frac{1}{n!}
\theta^{\mu_1\nu_1}\dots\theta^{\mu_n\nu_n}(\partial_{\mu_1}\dots\partial_{\mu_n}\xi_r^{\rho_1\dots\rho_r})
\partial_{\nu_1}\dots\partial_{\nu_n}\partial_{\rho_1\dots\rho_r} .\label{43}
\end{eqnarray}
where $f$ and $\xi_r$ are the images of $\hat{f}$ and $\hat{\xi}_r$ under $\varphi$:
\begin{equation}
\varphi: \quad \hat{f}\mapsto f {\mbox{ and }} \hat{\xi}_r\mapsto \xi _r .\label{44}
\end{equation}
Our aim is to study algebraic structures of the higher order differential
operators.

As an example we consider the deformed Lorentz transformations \cite{13}:
\begin{equation}
\hat{\delta}_\omega =-\hat{x}^\nu\omega_\nu^{\ \rho}\hp_\rho +\frac{i}{2}
\theta^{\rho\mu}\omega_\rho^{\ \nu}\hp_\mu\hp_\nu .\label{45}
\end{equation}
They represent the Lorentz algebra.
\begin{eqnarray}
\lb \delta_\omega, \delta_\omega' \rb &=& \delta_{\>\omega\times\omega'}, \nonumber\\
(\omega\times\omega')_\mu^{\ \nu} &=& -\big( \omega_\mu^{\ \sigma}\omega_\sigma^{'\ \nu} -
\omega_\mu^{'\ \sigma}\omega_\sigma^{\ \nu} \big). \label{46}
\end{eqnarray}
If we map the differential operator $\hat{\delta}_\omega$ to the commuting variables, we
obtain:
\begin{equation}
\varphi :\quad \hat{\delta}_\omega\mapsto -x^\nu\omega_\nu^{\ \rho}\partial_\rho .\label{47}
\end{equation}
This is the ``angular momentum'' representation of Lorentz transformations.
It is obvious that they satisfy the Lorentz algebra (\ref{46}), and this is the
reason why the operators (\ref{45}) satisfy (\ref{46}) as well.

\section{The $\star$-product of higher order differential operators}

In the previous section, in equation (\ref{43}), we introduced the $\star$-product of
a differential operator with a function:
\begin{equation}
X\star f=\Xi _X f. \label{48}
\end{equation}
This is a map of higher order differential operators
\begin{equation}
X\mapsto \Xi_X .\label{49}
\end{equation}
We want to show that this map is invertible. Given a differential
operator $\Xi$ we can find a differential operator $X_\Xi$ such that
\begin{equation}
\Xi f=X_\Xi\star f.\label{50}
\end{equation}
This is the inverse map
\begin{equation}
\Xi \mapsto X_\Xi .\label{51}
\end{equation}
We first define the differential operator $\Xi$:
\begin{equation}
\Xi = \sum _r \xi_r^{\lambda_1\dots\lambda_r}
\partial_{\lambda_1}\dots\partial_{\lambda_r} .\label{52}
\end{equation}
For the proof we use the tensor product notation and the identity
\begin{equation}
\mu\Big( e^{\frac{i}{2}\theta^{\mu\nu}\partial _\mu
\otimes\partial_\nu}
e^{-\frac{i}{2}\theta^{\rho\sigma}\partial _\rho
\otimes\partial_\sigma} \sum_r\xi_r^{\lambda_1\dots\lambda_r}\otimes
\partial_{\lambda_1}\dots\partial_{\lambda_r}f\Big) =\Xi f. \label{53}
\end{equation}
We now separate the $\star$-product:
{\begin{small}
\begin{eqnarray}
\Xi f\hspace*{-3mm}&=&\hspace*{-3mm}\mu\Big( e^{\frac{i}{2}\theta^{\mu\nu}\partial _\mu
\otimes\partial_\nu}
\sum_r\sum_n
\Big(-\frac{i}{2}\Big)^n\frac{1}{n!}
\theta^{\rho_1\sigma_1}\dots\theta^{\rho_n\sigma_n}(\partial_{\rho_1}\dots\partial_{\rho_n}
\xi_r^{\lambda_1\dots\lambda_r})\otimes
\partial_{\sigma_1}\dots\partial_{\sigma_n}\partial_{\lambda_1}\dots\partial_{\lambda_r}f \nonumber\\
&=&\hspace*{-3mm}
\sum_r\sum_n
\Big(-\frac{i}{2}\Big)^n\frac{1}{n!}
\theta^{\rho_1\sigma_1}\dots\theta^{\rho_n\sigma_n}(\partial_{\rho_1}\dots\partial_{\rho_n}
\xi_r^{\lambda_1\dots\lambda_r})\star
\partial_{\sigma_1}\dots\partial_{\sigma_n}\partial_{\lambda_1}\dots\partial_{\lambda_r}f \label{54}\\
&=&\hspace*{-3mm}
\sum_r\sum_n
\Big(-\frac{i}{2}\Big)^n\frac{1}{n!}
\theta^{\rho_1\sigma_1}\dots\theta^{\rho_n\sigma_n}(\partial_{\rho_1}\dots\partial_{\rho_n}
\xi_r^{\lambda_1\dots\lambda_r})
\partial_{\sigma_1}\dots\partial_{\sigma_n}\partial_{\lambda_1}\dots\partial_{\lambda_r}\star f .\nonumber
\end{eqnarray}
\end{small}}
We have found the differential operator $X_\Xi$
\begin{equation}
X_\Xi=\sum_r\sum_n
\Big(-\frac{i}{2}\Big)^n\frac{1}{n!}
\theta^{\rho_1\sigma_1}\dots\theta^{\rho_n\sigma_n}(\partial_{\rho_1}\dots\partial_{\rho_n}
\xi_r^{\lambda_1\dots\lambda_r})
\partial_{\sigma_1}\dots\partial_{\sigma_n}\partial_{\lambda_1}\dots\partial_{\lambda_r} .\label{55}
\end{equation}

This formula can be applied to the product of two functions
\begin{equation}
gf=\sum_n\Big(-\frac{i}{2}\Big)^n\frac{1}{n!}
\theta^{\rho_1\sigma_1}\dots\theta^{\rho_n\sigma_n}(\partial_{\rho_1}\dots\partial_{\rho_n}g)
\partial_{\sigma_1}\dots\partial_{\sigma_n}\star f =X_g\star f .\label{56}
\end{equation}
It immediately follows from (\ref{50}) that
\begin{equation}
X_g\star X_h\star f =g(X_h\star f)=ghf .\label{57}
\end{equation}
As a consequence, the differential operators $X_g$ and $X_h$
commute. This result can be used to define derivative dependent gauge
transformations in the $\star$-product formalism
\begin{equation}
\alpha(x)\psi(x)=X_\alpha\star\psi (x). \label{58}
\end{equation}

When applied to vector fields, formula (\ref{55}) allowes a deformation of the diffeomorphisms algebra.
\begin{equation}
\Xi_\xi=\xi^\mu\partial_\mu .\label{59}
\end{equation}
The corresponding differential operator $X_\xi$ is:
\begin{equation}
X_\xi=\sum_n\Big(-\frac{i}{2}\Big)^n\frac{1}{n!}
\theta^{\rho_1\sigma_1}\dots\theta^{\rho_n\sigma_n}(\partial_{\rho_1}\dots\partial_{\rho_n}\xi^\mu)
\partial_{\sigma_1}\dots\partial_{\sigma_n}\partial_\mu .\label{60}
\end{equation}
When we calculate the commutator of two such operators, we obtain
\begin{equation}
\lb X_\xi \ds X_\eta\rb =X_{\lb \eta,\> \xi\rb }, \label{61}
\end{equation}
where $\lb \eta, \xi\rb = (\eta^\mu\partial_\mu\xi^\nu-\xi^\mu\partial_\mu\eta^\nu)\partial_\nu$.
The differential operators $X$ have the Lie algebra structure of vector
fields.

Vector fields of the form
\begin{equation}
\Xi _\omega =x^\nu\omega_\nu^{\ \mu}\partial_\mu \label{62}
\end{equation}
form a finite-dimensional Lie algebra:
\begin{equation}
\lb \Xi_\omega,\Xi_{\omega '} \rb = x^\nu\lb \omega,\omega '\rb _\nu^{\ \mu}\partial_\mu ,\label{63}
\end{equation}
where $\lb \omega,\omega '\rb $ is the commutator of the matrices $\omega_\nu^{\ \mu}$.
The corresponding differential operators form the same algebra
\begin{equation}
\lb X_\omega \ds X_{\omega '}\rb =X_{\lb \omega ',\omega \rb }. \label{64}
\end{equation}

We have found the $\star$-product realization of the algebra of diffeomorphisms.
The Lorentz algebra (\ref{45}) can be obtained this way.

\section{The deformed algebra of diffeomorphisms and gauge transformations}

\subsection{Diffeomorphisms}

By equations (\ref{50}) and (\ref{61}) we have lifted the algebra of diffeomorphisms to the $\star$-product
realization that, in turn, leads to a quantum space realization by the inverse homeomorphism  $\varphi ^{-1}$.
To study the comultiplication we apply the differential operator $X_\xi$ to the $\star$-product
of two functions. By the very definition of $X_\xi$ we obtain:
\begin{equation}
X_\xi\star f\star g=\xi^\rho\partial_\rho(f\star g) .\label{65}
\end{equation}
The parameter $\xi(x)$ does not commute with the derivatives of the $\star$-product,
additional terms arise. We calculate to first order in $\theta$:
\begin{equation}
\xi^\rho\partial_\rho(f\star g)=(\xi^\rho\partial_\rho f)\star g + f\star (\xi^\rho\partial_\rho g)
+\frac{i}{2}\Big( \theta^{\beta\nu}(\partial_\nu \xi^\alpha) - \theta^{\alpha\nu}(\partial_\nu \xi^\beta)\Big)
(\partial_\alpha f)(\partial _\beta g) .\label{66}
\end{equation}
To first order in $\theta$ this can be summarized in the comultiplication rule:
\begin{equation}
\Delta(X_\xi)=X_\xi\otimes 1+1\otimes X_\xi +\frac{i}{2}\theta^{\alpha\nu}
\Big( (\partial_\nu X_\xi)\otimes\partial_\alpha - \partial_\alpha \otimes\partial_\nu X_\xi\Big) .\label{67}
\end{equation}
This is the new comultiplication law of the bialgebra of diffeomorphisms to first order in $\theta$.
It needs some calculation to show that this comultiplication law is compatible with the algebra:
\begin{equation}
\lb \Delta(X_\xi) \ds \Delta(X_\eta) \rb = \Delta(X_{\lb \eta,\xi\rb }) .\label{68}
\end{equation}
This comultiplication rule can be calculated to all orders in $\theta$, the result will be published in a forthcoming paper.
Equations (\ref{61}), (\ref{67}) and (\ref{68}) allow us to treat the deformed Lie algebra of
diffeomorphisms on a purely formal level.

Again, the algebra remains the same, the comultiplication is different, this yields a deformed bialgebra of diffeomorphisms.

\subsection{Gauge transformations}

Derivative valued gauge transformations have been defined. The parameter $\alpha$ will be Lie algebra valued.
At the classical level we have:
\begin{equation}
\delta_\alpha \psi(x) = i\alpha(x)\psi(x) = i\alpha _b T^{b}\psi(x) .\label{69}
\end{equation}
The matrices $T^a$ are the generators of the Lie algebra in the respective representation spanned by $\psi$.
This can be lifted to a $\star$-product realization:
\begin{equation}
\hat{\delta}_\alpha\star \psi = iX_\alpha\star\psi = i \alpha \psi .\label{70}
\end{equation}
This equation defines the differential operator $X_\alpha$.
To first order in $\theta$ we obtain with Lie algebra valued $\alpha$ :
\begin{equation}
X_\alpha=\alpha-\frac{i}{2}\theta^{\mu\nu}(\partial_\mu\alpha )\partial_\nu .\label{71}
\end{equation}

We calculate the commutator of two such transformations:
\begin{eqnarray}
\hat{\delta}_\beta\star\hat{\delta}_\alpha\star \psi\hspace*{-3mm}&=&\hspace*{-3mm}-X_\alpha\star X_\beta\star\psi
= - \alpha\beta \nonumber\\
\lb \hat{\delta}_\beta\ds\hat{\delta}_\alpha \rb \hspace*{-3mm}&=&\hspace*{-3mm}\lb \beta,\alpha\rb \psi =
\hat{\delta}_{\lb \beta,\alpha\rb }\psi \label{72}
\end{eqnarray}
and we see that they form a realization of the gauge group.

The comultiplication can be derived from the application of $\hat{\delta}_\alpha$ to the $\star$-product of
two fields $\psi$ and $\phi$:
\begin{eqnarray}
X_\alpha\star\psi\star\phi\hspace*{-3mm}&=&\hspace*{-3mm}\alpha(\psi\star\phi)
=\alpha_b \Big( T^b\psi\star\phi + \psi\star T^b\phi \Big) \label{73} \\
&=&\hspace*{-3mm}(\alpha_b T^b\psi)\star\phi + \psi\star (\alpha_b T^b\phi)
-\frac{i}{2}\theta^{\mu\nu}\Big( (\partial_\mu \alpha_b)T^b\psi\star\partial_\nu\phi
+(\partial_\mu \psi)\star (\partial_\nu\alpha_b) T^b\phi \Big)\nonumber
\end{eqnarray}
all to first order in $\theta$. We abstract the comultiplication rule:
\begin{equation}
\Delta(X_\alpha)=X_\alpha\otimes 1+1\otimes X_\alpha -\frac{i}{2}\theta^{\mu\nu}
\Big( (\partial_\mu X_\alpha)\otimes\partial_\nu + \partial_\mu\otimes\partial_\nu X_\alpha\Big) .\label{74}
\end{equation}
It again needs some calculation to verify directly
\begin{equation}
\lb \Delta(X_\alpha) \ds \Delta(X_\beta) \rb = \Delta(\lb X_\beta \ds X_\alpha\rb ) .\label{75}
\end{equation}

\section{Fields}

Fields are elements of $\hat{\cal{A}}_{\hat{x}}$ with certain transformation properties.
We formulate these properties in the $\star$-realization of the algebra $\hat{\cal{A}}_{\hat{x}}$.
For a scalar field we define:
\begin{equation}
\delta_\xi\phi = -X_\xi\star\phi .\label{76}
\end{equation}
The differential operator $X_\xi$ was defined in (\ref{60}). It is
\begin{eqnarray}
X_\xi\hspace*{-3mm}&=&\hspace*{-3mm}\sum_n\Big(-\frac{i}{2}\Big)^n\frac{1}{n!}
\theta^{\rho_1\sigma_1}\dots\theta^{\rho_n\sigma_n}(\partial_{\rho_1}\dots\partial_{\rho_n}\xi^\mu)
\partial_{\sigma_1}\dots\partial_{\sigma_n}\partial_\mu \nonumber\\
&=&\hspace*{-3mm} \xi^\mu\partial_\mu -\frac{i}{2}\theta^{\rho\sigma}\partial_\rho\xi^\mu\partial_\nu\partial_\mu\dots .
\label{77}
\end{eqnarray}
It has the property:
\begin{equation}
X_\xi\star\phi =\xi^\mu\partial_\mu\phi .\label{78}
\end{equation}

With the definition (\ref{76}) we have lifted the diffeomorphism algebra that generates
the general coordinate transformations in ${\cal{A}}_x$ to a realization on $\hat{\cal{A}}_{\hat{x}}$.
We have seen in the previous section that the comultiplication changes.

Finally, we can lift the concept of vector fields and tensor fields to $\hat{\cal{A}}_{\hat{x}}$ as well.
A covariant vector field in ${\cal{A}}_x$ has the transformation properties:
\begin{equation}
\delta_\xi V_\mu = -\xi^\rho\partial_\rho V_\mu -(\partial _\mu \xi^\rho)V_\rho .\label{79}
\end{equation}
It agrees with the transformation of the derivative of a scalar field.
\begin{equation}
\delta_\xi \partial_\mu \phi = -\xi^\rho\partial_\rho \partial_\mu \phi
-(\partial _\mu \xi^\rho)\partial_\rho \phi .\label{80}
\end{equation}
Equation (\ref{79}) can be lifted to $\hat{\cal{A}}_{\hat{x}}$
\begin{equation}
\delta_\xi V_\mu = -X_\xi \star V_\mu -X_\mu^{\ \rho}\star V_\rho .\label{81}
\end{equation}
where $X_\mu^{\ \rho}$ is the differential operator
\begin{equation}
X_\mu^{\ \rho} = (\partial _\mu \xi^\rho) -\frac{i}{2}\theta^{\sigma\nu}
(\partial_\sigma\partial_\mu \xi^\rho)\partial_\nu +\dots \label{82}
\end{equation}
that can be computed from (\ref{79}).

The same procedure applies to contravariant vector fields as well as to tensor fields with an arbitrary number of indices.
Thus, we represent the diffeomorphism algebra on these vector fields.
For contravariant vector fields we have
\begin{equation}
\delta_\xi V^\mu = -\xi^\rho\partial_\rho V^\mu +(\partial _\rho \xi^\mu)V^\rho .\label{83}
\end{equation}
This transformation can be lifted to the $\star$-product realisation
\begin{equation}
\delta_\xi V^\mu = -X_\xi \star V^\mu +X^\mu_{\ \rho}\star V^\rho ,\label{84}
\end{equation}
with the differential operator
\begin{equation}
X^\mu_{\ \rho} = (\partial _\rho \xi^\mu) -\frac{i}{2}\theta^{\sigma\nu}
(\partial_\sigma\partial_\rho \xi^\mu)\partial_\nu +\dots .\label{85}
\end{equation}

\end{document}